\documentclass[12pt]{article}
\usepackage{amsmath}
\usepackage{graphicx,psfrag,epsf}
\usepackage{enumerate}
\usepackage{natbib}
\usepackage{url} 
\usepackage{amsthm}
\usepackage{bbm}
\usepackage{mathtools}
\usepackage{amsfonts}
\usepackage{multirow, booktabs}
\usepackage[table]{xcolor}
\usepackage{cleveref}
\usepackage[T1]{fontenc}
\usepackage[utf8]{inputenc}
\usepackage{authblk}
\usepackage[multiple]{footmisc}
\usepackage{blindtext,titlefoot}
\usepackage{sectsty}
\usepackage{graphicx,subfigure}

\usepackage{amsmath}
\usepackage{amssymb}
\usepackage{amsfonts}
\usepackage{multirow}
\usepackage{amsthm}
\usepackage{url}
\usepackage{xcolor}
\usepackage{float}
\usepackage{algorithm}
\usepackage{algorithmic}
\usepackage{rotating}
 
\floatname{algorithm}{Procedure}

\newcommand{\bH}{\mathbf{H}}

\newcommand{\bQ}{{\boldsymbol{Q}}}
\newcommand{\bI}{{\boldsymbol{\mathcal{I}}}}

\newtheorem{theorem}{Theorem}

\theoremstyle{definition}

\newtheorem{remark}{Remark}
\newtheorem{condition}{Condition}
\newtheorem{Proposition}{Proposition}

\makeatletter
\newcommand*{\rom}[1]{\expandafter\@slowromancap\romannumeral #1@}
\makeatother

\begin{document}

\sectionfont{\bfseries\large\sffamily}%
%

\subsectionfont{\bfseries\sffamily\normalsize}%
%


\def\spacingset#1{\renewcommand{\baselinestretch}%
{#1}\small\normalsize} \spacingset{1}

\title{\Large \bf  Bias Mitigation in Matched Observational Studies with Continuous Treatments: Calipered Non-Bipartite Matching and Bias-Corrected Estimation and Inference}
\date{}

 \author{Anthony Frazier\thanks{Anthony.Frazier@colostate.edu}\vspace{.1cm}\\
   Department of Statistics, Colorado State University\vspace{.3cm}\\ 
   Siyu Heng\thanks{siyuheng@nyu.edu}\vspace{.1cm}\\ 
    Department of Biostatistics, School of Global Public Health, \\New York University\vspace{.3cm}\\ 
    Wen Zhou \thanks{w.zhou@nyu.edu}\vspace{.1cm}\\
     Department of Biostatistics, School of Global Public Health, \\New York University
    }

\maketitle

\begin{abstract}
In matched observational studies with continuous treatments, individuals with different treatment doses but the same or similar covariate values are paired for causal inference. While inexact covariate matching (i.e., covariate imbalance after matching) is common in practice, previous matched studies with continuous treatments have often overlooked this issue as long as post-matching covariate balance meets certain criteria. Through re-analyzing a matched observational study on the social distancing effect on COVID-19 case counts, we show that this routine practice can introduce severe bias for causal inference. Motivated by this finding, we propose a general framework for mitigating bias due to inexact matching in matched observational studies with continuous treatments, covering the matching, estimation, and inference stages. In the matching stage, we propose a carefully designed caliper that incorporates both covariate and treatment dose information to improve matching for downstream treatment effect estimation and inference. For the estimation and inference, we introduce a bias-corrected Neyman estimator paired with a corresponding bias-corrected variance estimator. The effectiveness of our proposed framework is demonstrated through numerical studies and a re-analysis of the aforementioned observational study on the effect of social distancing on COVID-19 case counts. An open-source $\texttt{R}$ package for implementing our framework has also been developed.
\end{abstract}

\spacingset{1.75}

\section{Introduction}
\label{sec:intro}
Matching is a widely used causal inference framework in observational studies for both binary treatments, where it is known as \textit{bipartite matching}, and continuous treatments, where it is referred to as \textit{non-bipartite matching} \citep{rubin1979using, rosenbaum2002observational, lu2001matching, baiocchi2010building, zubizarreta2013stronger, yang2014dissonant, greevy2023optimal, zhang2023statistical, zhang2024sensitivity}. It aims to mimic a randomized experiment by pairing individuals with different treatment values but same or similar covariates. Under exact matching, where covariates are exactly matched within pairs, treatments are as-if randomly assigned within each pair, allowing direct application of conventional randomization-based inference. For example, the conventional Neyman (difference-in-means) estimator is an unbiased estimator of the sample average treatment effect (SATE) under exact matching, and an asymptotically valid confidence interval for the SATE can be constructed using the Neyman-type variance estimator \citep{rosenbaum2002observational, baiocchi2010building}. 

However, matching is often inexact in practice, especially when continuous covariates and/or many covariates are involved. Previous observational studies have routinely treated inexactly matched datasets as if they were exactly matched, applying randomization-based inference as long as covariate balance assessments were met \citep{rosenbaum2002observational, rosenbaum2020design, gagnon2019classification, branson2021randomization, chen2023testing, zhang2024sensitivity}. In the binary treatment case, recent studies have shown that ignoring inexact matching can introduce severe bias in downstream randomization-based inference and have proposed strategies to address this issue \citep{guo2023statistical, pimentel2024covariate, zhu2023bias}. However, to our knowledge, an effective strategy for mitigating bias due to inexact matching \textit{in the continuous treatment case} remains largely unexplored.

\subsection{Motivation: Social Distancing Effects in a Matched Observational COVID-19 Study} \label{subsec: motivating example}

\begin{figure}[H]
 \caption{The absolute standardized difference-in-means for all covariates among the $1,211$ matched pairs of U.S. counties formed by \cite{zhang2023social}. All covariates have an absolute standardized difference-in-means below $0.2$, indicated by the dashed line.}
    \centering
    \includegraphics[width=0.8\linewidth]{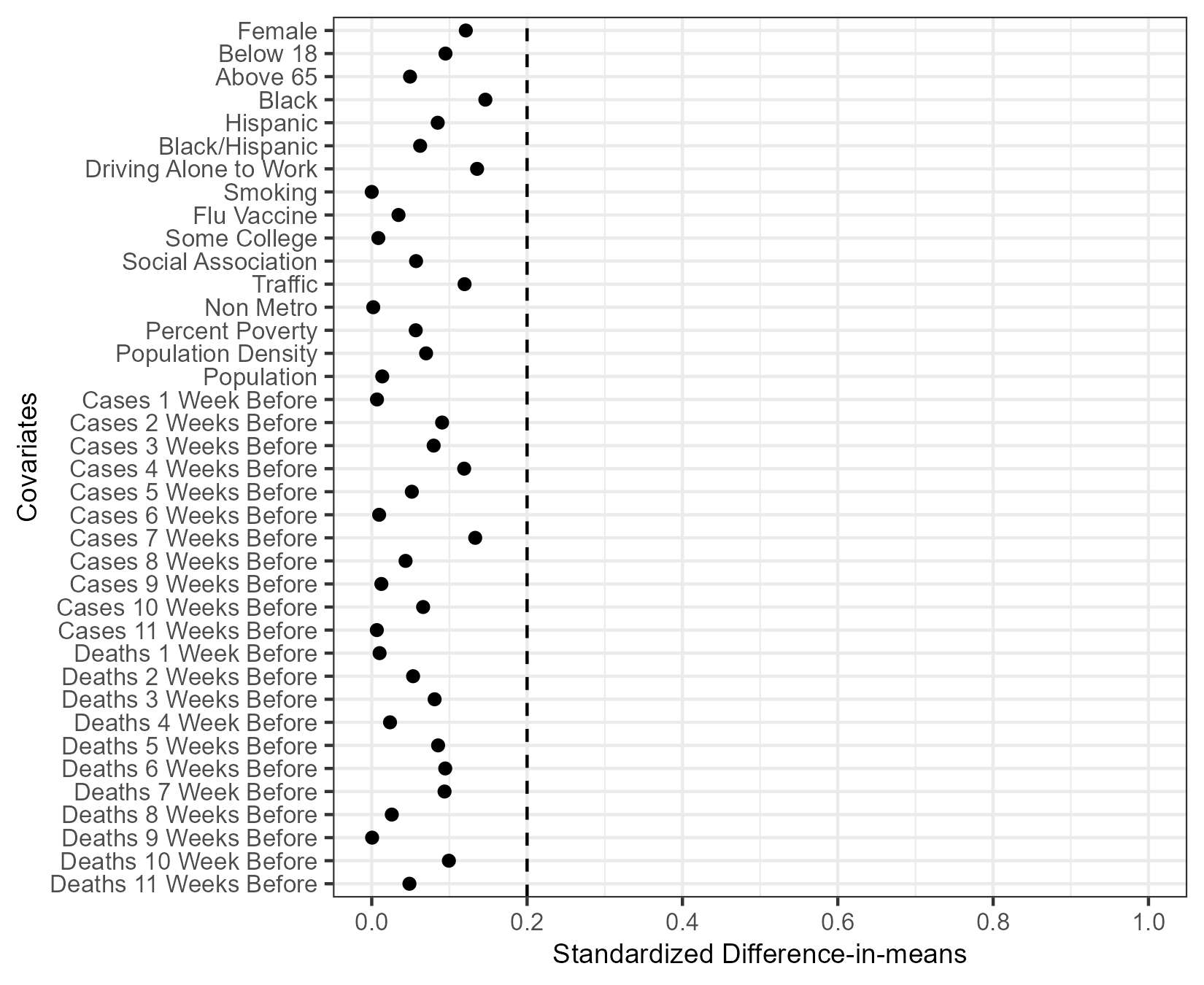}
    \label{fig:covariate_balance}
\end{figure}

Several early studies on the COVID-19 pandemic identified associations between social distancing practices and reduced COVID-19 cases \citep{lewnard2020scientific, lau2020positive, sjodin2020only, keller2022tracking}. To examine whether these associations are causal, we re-analyze the COVID-19 and social distancing data studied by \cite{zhang2023social} which integrates datasets from multiple sources, including \texttt{Unacast}\texttrademark, the United States Census Bureau, and the County Health Rankings and Roadmaps Program. Like \cite{zhang2023social}, we focus on the county-level causal relationship between social distancing and cumulative COVID-19 cases in the United States. The treatment dose is a continuous county-level social distancing measure during the re-opening phase (April 27th to June 28th, 2020), defined as the average percentage change in total distance traveled during this period compared to the baseline before the pandemic. For example, a value of $-0.1$ or $+0.1$ indicates a $10\%$ decrease or increase in total distance traveled. The outcome variable is the cumulative COVID-19 case count per $100,000$ people during the follow-up period from June 29th to August 2nd, 2020. To study the effect of social distancing, \citet{zhang2023social} formed $1,211$ matched pairs of U.S. counties with different social distancing scores but similar covariate values, including county-level sociodemographic and socioeconomic measures, as well as baseline COVID-19 case counts before the re-opening phase (see Table~\ref{tab: covariate description} for details). We use the same $1,211$ matched pairs, with covariate balance reported in Figure~\ref{fig:covariate_balance}. The absolute standardized differences in means for all the considered covariates between paired counties are below $0.15$, within the commonly accepted threshold of $0.2$, which \cite{zhang2023social} considered evidence of adequate covariate balance.

\begin{table}
\centering
\caption{Description of each covariate considered in the data analysis.}
\centering
\begin{tabular}{ll}
\toprule
Covariate & Description \\
\midrule
Female (\%) & Population identified as female \\
Below 18 (\%) & Population below the age of 18 \\
Above 65 (\%) & Population above the age of 18 \\
Black/Hispanic (\%) & Population identified as either black or Hispanic \\
Driving Alone to Work (\%) & Population that drives alone to work \\
Smoking (\%) & Adult population that has been identified as current smokers \\
\multirow{2}{*}{Flu Vaccine (\%)} & Fee-For-Service (FFS) Medicare enrollees that had \\
& an annual flu vaccination \\
Some College (\%) & Population with some college education \\
Social Association &  Number of membership associations per 10,000 people \\
\multirow{2}{*}{Non-Metro (0/1)} & Equals $1$ if county is ``rural" \\
& (determined by the Rural-Urban Continuum Codes) \\
Percent Poverty (\%) & Population below the poverty line in 2018 \\
Population Density & Average population per square mile \\
Population & Total population size of the county \\
\multirow{2}{*}{Cases ... Weeks Before} & Count of COVID-19 cases ... weeks before \\
& June 29th 2020 \\
\multirow{2}{*}{Deaths ... Weeks Before} & Count of COVID-19-related deaths ... weeks \\
& before June 29th 2020 \\
\bottomrule
\end{tabular}
\label{tab: covariate description}
\end{table}

In examining the causal relationship between social distancing and COVID-19 cases, \cite{zhang2023social} considers testing Fisher's sharp null of no treatment effect for each county and conducts a secondary dose-response analysis using a parametric kink model. In our re-analysis, we focus on the SATE, defined as the sample mean difference between potential outcomes under paired treatment doses, normalized by the mean difference in paired treatment doses \citep{baiocchi2010building, rosenbaum2020design}. Using the conventional Neyman difference-in-means estimator, our SATE estimate for the effect of social distancing on COVID-19 case counts is $-1041.478$ with a $95\%$ confidence interval of $(-1399.777, -683.180)$. This suggests strong evidence that stricter social distancing policies are associated with an \textit{increase} in COVID-19 case counts, which is contrary to prior studies on the effects of social distancing and lockdown measures \citep{lewnard2020scientific, lau2020positive, sjodin2020only}. This indicates that the routine practice of naively treating inexactly matched datasets as exactly matched during outcome analysis may severely bias causal inference results.

\subsection{Our Contributions}\label{subsec: our contributions}

In this work, we propose a general framework for bias mitigation in inexactly matched observational studies with continuous treatments, covering the matching, estimation, and inference stages. For the matching stage, we introduce the \textit{dose assignment discrepancy caliper}, a new type of caliper that uses estimated post-matching treatment dose assignment probabilities. By leveraging both covariate and treatment dose information, it prevents forming matched pairs with estimated post-matching treatment dose assignment probabilities substantially deviant from random assignments. To our knowledge, this is the first caliper in the non-bipartite matching literature that fully utilizes treatment dose information without discretizing the continuous treatment variable. For the estimation stage, we propose a novel \textit{bias-corrected Neyman estimator} that corrects bias incurred from inexact covariate matching, which is the first bias-corrected estimator for the SATE beyond the binary treatment case. For the inference stage, we derive a valid variance estimator for the proposed bias-corrected Neyman estimator, hereafter referred to as the \textit{bias-corrected variance estimator}. Our variance estimator is the first in the non-bipartite matching literature to be (asymptotically) valid under inexact matching and heterogeneous treatment effects.

Our proposed methods -- the new caliper for non-bipartite matching and the bias-corrected estimation and inference procedure -- can be applied either separately or in combination towards continuous, discrete, and ordinal treatment cases. Simulation studies demonstrate that our proposed methods can substantially reduce estimation bias and improve the coverage rates of confidence intervals for the SATE. As an important data application, we re-analyze the motivating example of social distancing's effect on COVID-19 case counts in the United States using our proposed methods, revealing interesting causal findings that align with early scientific intuitions. We have also developed an open-source $\texttt{R}$ package \texttt{nbpInference} (available at \url{https://github.com/AnthonyFrazierCSU/nbpInference}) for implementation of our proposed methods.

\section{Review of the Common Practice for Matched Observational Studies with Continuous Treatments}\label{sec: review}

\subsection{Review of the Vanilla Non-Bipartite Matching Framework}

In matched observational studies with continuous treatments, statistical non-bipartite matching is a common framework for forming matched pairs or sets \citep{lu2001matching, lu2011optimal, greevy2023optimal, zhang2023statistical}. This framework pairs individuals with different treatment doses but same or similar covariate values, typically by optimizing the allocation of matches to minimize some total distances between matched individuals' covariates, such as Mahalanobis distances, sometimes weighted by treatment dose differences. Below, we describe non-bipartite matching under a graph-theoretic framework, where we focus on the setting of optimal pair matching without replacement, one of the most fundamental and widely used non-bipartite matching methods (\citealp{lu2001matching, lu2011optimal, baiocchi2010building, rosenbaum2020modern, heng2023instrumental, zhang2023social}). For other non-bipartite matching designs, see \citet{wu2024matching}, \citet{zhang2023statistical}, and \citet{zhang2024bridging}.

Suppose that there are $N$ individuals. Consider a graph consisting of a pair of sets $(V, E)$, where $V = \{v_{n}: n=1,\dots, N\}$ represents the set of nodes, and $E = \{(v_{n_{1}}, v_{n_{2}}): v_{n_{1}}, v_{n_{2}} \in V, n_{1}<n_{2} \}$ represents the set of edges, connecting pairs of nodes. In this setting, let $V$ represent the study sample before matching, with each node corresponding to an individual, and $E$ represents the set of matched pairs formed from $V$. In the optimal non-bipartite matching framework, there is no predefined treatment or control group, and the only constraints for edges in $E$ are: 1) $(v_{n}, v_{n})\notin E$ for all $n=1,\dots, N$, preventing an individual from being matched with itself; and 2) if $(v_{n_{1}}, v_{n_{2}})\in E$ for some $n_{1}$ and $n_{2}$, then for any $n_{3} \neq n_{2}$, we have $(v_{n_{1}}, v_{n_{3}})\notin E$, which indicates matching without replacement.  Let $\mathcal{E}$ denote the collection of all $E$ that satisfy these constraints. The optimal set of matched pairs $E^{*} \in \mathcal{E}$ minimizes the total distances between matched individuals. Formally,
\begin{equation}\label{eqn: nbp matching}
    E^{*} = \underset{E \in \mathcal{E}}{\text{argmin}}\sum_{(v_{n_{1}}, v_{n_{2}})\in E} d(v_{n_{1}}, v_{n_{2}}),
\end{equation}
where $d(v_{n_{1}}, v_{n_{2}})$ is some prespecified distance metric between nodes $v_{n_{1}}$ and $v_{n_{2}}$. One widely used $d(\cdot,\cdot)$ is the Mahalanobis distance between the covariates of individuals $n_{1}$ and $n_{2}$ (\citealp{rosenbaum1985constructing, lu2001matching, lu2011optimal}). We refer to \citet{lu2001matching, lu2011optimal} and \citet{greevy2023optimal} for other commonly used distance metrics.

\subsection{Review of Randomization-Based Estimation and Inference Assuming Exact Matching}\label{subsec: review of randomization inference assuming exact matching}

Now, we review the commonly used randomization-based estimation and inference framework assuming exact non-bipartite matching. That is, one assumes that the covariates are exactly matched within each matched pair. Suppose there are $I$ matched pairs formed by non-bipartite matching. For individual $j\in\{1,2\}$ in matched pair $i$, let $Z_{ij}$ be the observed treatment dose, and let $\mathbf{x}_{ij}=(x_{ij1}, \dots, x_{ijK})$ denote the $K$-dimensional observed covariates. In an exactly matched study, the two individuals in each pair have identical covariate values, where $\mathbf{x}_{i1}=\mathbf{x}_{i2}$, but differ in treatment doses, meaning $Z_{i1}\neq Z_{i2}$ \citep{rosenbaum1989sensitivity, gastwirth1998dual, lu2001matching}. For paired individuals $i1$ and $i2$, the lower and higher treatment dose are denoted as $z_{i}^{*}=Z_{i1}\wedge Z_{i2}$ and $z^{**}_{i}=Z_{i1}\vee Z_{i2}$, respectively. Under exact matching and the ignorability assumption \citep{rosenbaum1983central, rosenbaum2002observational}, the two paired individuals have equal probabilities of receiving either the higher or lower treatment dose. That is,   
\begin{align}\label{eqn: random assign}
    &\quad \mathbb{P}(Z_{i1}=z_{i}^{*}, Z_{i2}=z^{**}_{i}| \mathbf{x}_{i1}=\mathbf{x}_{i2}, Z_{i1}\wedge Z_{i2}=z_{i}^{*}, Z_{i1}\vee Z_{i2}=z_{i}^{**})\nonumber \\
    &= \mathbb{P}(Z_{i1}=z^{**}_{i}, Z_{i2}=z_{i}^{*}| \mathbf{x}_{i1}=\mathbf{x}_{i2}, Z_{i1}\wedge Z_{i2}=z_{i}^{*}, Z_{i1}\vee Z_{i2}=z_{i}^{**})=1/2. 
\end{align}
Let $Y_{ij}$ denote the observed outcome for individual $j$ in matched pair $i$. Following the potential outcomes framework \citep{neyman1923application, rubin1974estimating, rosenbaum1989sensitivity}, let $Y_{ij}(z)$ denote the potential outcome of individual $ij$ under the treatment dose $Z_{ij}=z$. In a randomization-based estimation and inference framework, the potential outcomes $Y_{ij}(z)$ are treated as fixed values, while the randomness for inference is due to the random assignment of treatment doses after matching, as specified in \eqref{eqn: random assign} \citep{rosenbaum1989sensitivity, rosenbaum2020design, gastwirth1998dual, baiocchi2010building, zhang2023social}. 

Define $\mathbf{Z}=(Z_{11}, \dots, Z_{I2})$ as the vector of treatment doses and $\mathbf{Y}=(Y_{11}, \dots, Y_{I2})$ as the vector of observed outcomes. Let $\mathcal{Z}=\{\mathbf{Z}: (Z_{i1}, Z_{i2})=(z_{i}^{*}, z_{i}^{**}) \text{ or } (Z_{i1}, Z_{i2})=(z_{i}^{**}, z_{i}^{*})\}$ denote all possible treatment dose assignments conditional on matching, where $|\mathcal{Z}|=2^{I}$. Define $\mathcal{F}=\{(Y_{ij}(z_{i}^{*}), Y_{ij}(z^{**}_{i}), \mathbf{x}_{ij}): i=1,\dots, I, j=1,2\}$ as the collection of potential outcomes for matched pairs and observed covariates for all individuals. Denote by $\widetilde{Z}_{ij}=\mathbbm{1}\{Z_{ij}=z_{i}^{**}\}$ the indicator of whether individual $j$ in matched pair $i$ receives the higher dose $z_{i}^{**}$, ensuring $\widetilde{Z}_{i1}+\widetilde{Z}_{i2}=1$ for each $i$.

In matched observational studies with continuous treatments, a widely considered causal estimand is the (generalized) sample average treatment effect (SATE) $\lambda$, defined as the ratio of the total post-matching treatment effects to the total paired differences in treatment doses:
\begin{equation*}
     \lambda=\frac{\sum_{i=1}^{I}\sum_{j=1}^{2}\{ Y_{ij}(z^{**}_{i})-Y_{ij}(z_{i}^{*})\}}{\sum_{i=1}^{I}\sum_{j=1}^{2}(z^{**}_{i}-z_{i}^{*})}
     =\frac{\sum_{i=1}^{I}\sum_{j=1}^{2}\{Y_{ij}(z^{**}_{i})-Y_{ij}(z_{i}^{*})\}}{2\sum_{i=1}^{I}(z^{**}_{i}-z_{i}^{*})},
\end{equation*}
which is also referred to as the \textit{effect ratio} in the literature \citep{baiocchi2010building, rosenbaum2020design}. When there are only two treatment doses and therefore $Z_{ij}$ is binary, $\lambda$ reduces to the classic SATE for binary treatments: $(2I)^{-1}\sum_{i=1}^{I}\sum_{j=1}^{2}\{ Y_{ij}(1)-Y_{ij}(0)\}$ \citep{neyman1923application, imbens2015causal, rigdon2015randomization,li2017general}. The commonly used estimator for $\lambda$ is the (generalized) Neyman estimator \citep{baiocchi2010building, zhang2022bridging, heng2023instrumental}, given by 
\begin{equation*}
    \widehat{\lambda}_{N}=\frac{\sum_{i=1}^{I}(\widetilde{Z}_{i1}-\widetilde{Z}_{i2})(Y_{i1}-Y_{i2})}{\sum_{i=1}^{I}(z^{**}_{i}-z_{i}^{*})}.
\end{equation*}
When there are only two unique treatment doses, $\widehat{\lambda}_{N}$ reduces to the conventional Neyman estimator for binary treatments. Under exact matching with \eqref{eqn: random assign} holds, it is straightforward to show that $\mathbb{E}(\widehat{\lambda}_{N}| \mathcal{F}, \mathcal{Z})=\lambda$, so that the Neyman estimator $\widehat{\lambda}_{N}$ is unbiased to estimate the effect ratio $\lambda$. Also, assuming exact matching, the corresponding Neyman-type variance estimator for $\widehat{\lambda}_{N}$ can be derived using the arguments in \cite{baiocchi2010building}, allowing the construction of confidence intervals with asymptotic coverage guarantee.

\section{Our Proposed Methods for Mitigating Bias due to Inexact Non-Bipartite Matching}
 
In practice, non-bipartite matching typically cannot be exact, meaning $\mathbf{x}_{i1}=\mathbf{x}_{i2}$ may not hold for some or even any matched pair $i$, especially when dealing with continuous and/or many covariates. Previous studies have often overlooked inexact matching, directly applying classical randomization-based estimation and inference under the assumption of exact matching, as long as some empirical post-matching covariate balance criteria were satisfied, such as the absolute standardized mean differences being below $0.2$ for all covariates \citep{rosenbaum2020design, zhang2022bridging, zhang2023social, heng2023instrumental}. However, as demonstrated by the motivating example in Section~\ref{subsec: motivating example} and our simulation studies in Section~\ref{sec: simulation studies}, this common practice can introduce severe bias to the causal inference. To improve analysis of matched observational studies with continuous treatments, we propose a bias mitigation framework consisting of two key parts: {\it calipered non-bipartite matching} and {\it bias-corrected estimation and inference}.

\subsection{Part One: Tailoring Non-Bipartite Matching with A Dose Assignment Discrepancy Caliper}\label{sec: calipered matching}

In the bipartite matching literature (i.e., matching with binary treatments), researchers often use a propensity score caliper to reduce bias due to inexact covariate matching. This approach discourages or prevents individuals from being matched if their propensity scores, defined as the probability of receiving treatment conditional on covariates, differ substantially \citep{rosenbaum1985constructing, rosenbaum2020design, yu2020matching}. This can ensure that each matched pair or set achieves an acceptable matching quality. However, in the non-bipartite matching literature, a well-defined and sensible method for constructing such calipers remains lacking. Some recent studies circumvent complications with continuous treatments by dichotomizing the continuous treatment variable $Z$ into a binary variable $A$, assigning it a value of $1$ if the original treatment dose $Z$ exceeds the median and $0$ otherwise \citep{keele2020stronger, fogarty2021biased, yu2023risk}. Then, they design calipers based on the propensity score of this binary variable $A$, i.e., the probability of $A = 1$ given the covariates. Nevertheless, this strategy fails to make full use of the treatment dose information and is difficult to directly integrate into downstream estimation and inference, where the exact treatment dose should typically be considered.

In this section, we introduce a new type of caliper called the \textit{dose assignment discrepancy caliper}. Our new caliper is motivated by observing a key limitation of the commonly used distance metrics $d(v_{n_{1}}, v_{n_{2}})$ in \eqref{eqn: nbp matching} for non-bipartite matching: they fail to incorporate the explicit form of the post-matching treatment dose assignment probabilities that enter into randomization-based estimation and inference, i.e., the probability that individual $n_{1}$ or $n_{2}$ receives the higher treatment doses $z^{**}= Z_{n_{1}}\vee Z_{n_{2}}$ if they were paired: 
\begin{align*}
p_{n_{1}}&=\mathbb{P}(Z_{n_{1}}=z^{**}, Z_{n_{2}}=z^{*}| \mathbf{x}_{n_{1}}, \mathbf{x}_{n_{2}}, Z_{n_{1}}\wedge Z_{n_{2}}=z^{*}, Z_{n_{1}}\vee Z_{n_{2}}=z^{**}),~ \text{or}~\\ p_{n_{2}}&=\mathbb{P}(Z_{n_{1}}=z^{*}, Z_{n_{2}}=z^{**}| \mathbf{x}_{n_{1}}, \mathbf{x}_{n_{2}}, Z_{n_{1}}\wedge Z_{n_{2}}=z^{*}, Z_{n_{1}}\vee Z_{n_{2}}=z^{**}).\end{align*} Although the true values of these post-matching treatment dose assignment probabilities may be unknown, their estimates $\widehat{p}_{n_{1}}$ and $\widehat{p}_{n_{2}}$ can be leveraged in the matching stage to improve the vanilla distance metric $d(v_{n_{1}}, v_{n_{2}})$. Specifically, we propose adding a penalty term to the distance metric as follows:
\begin{equation}\label{eqn: distance metric with penalty}
    d^{*}(v_{n_{1}}, v_{n_{2}})=d(v_{n_{1}}, v_{n_{2}})+M\cdot \mathbbm{1}\left\{\min(\widehat{p}_{n_{1}}, \widehat{p}_{n_{2}})<\xi \right\},
\end{equation}
where $M$ is a sufficiently large number, and $\xi< 1/2$ is a small prespecified threshold (e.g., $\xi=0.1
$). If the estimated post-matching treatment dose assignment probability $\widehat{p}_{n_{1}}$, or equivalently, $\widehat{p}_{n_{2}}$, greatly departs from random assignments, i.e., diverges from $\widehat{p}_{n_{1}}=\widehat{p}_{n_{2}}=1/2$, we aim to avoid matching these individuals using \eqref{eqn: distance metric with penalty}. Therefore, we call the indicator $\mathbbm{1}\left\{\min(\widehat{p}_{n_{1}}, \widehat{p}_{n_{2}})<\xi \right\}$ as the \textit{dose assignment discrepancy caliper}. Compared to the vanilla distance metric $d(v_{n_{1}}, v_{n_{2}})$ without a caliper, the adjusted distance metric $d^{*}(v_{n_{1}}, v_{n_{2}})$ with the dose assignment discrepancy caliper is more favorable, as it prevents matching individuals with seemingly similar covariates, such as those with a small Mahalanobis distance, but highly non-random post-matching dose assignment probabilities. This issue arises, for instance, when two individuals are similar in most covariates except for a single key covariate, which precisely plays the most influential role in predicting the generalized propensity score $f(z | \mathbf{x})$ (i.e., the conditional density of treatment dose given the covariates) \citep{imai2004causal, wu2024matching}.
 
In practice, the estimated probabilities $\widehat{p}_{n_{1}}$ and $\widehat{p}_{n_{2}}$ used in the caliper $\mathbbm{1}\left\{\min(\widehat{p}_{n_{1}}, \widehat{p}_{n_{2}})<\xi \right\}$ can be obtained through the following process. First, assuming that individuals' treatment doses are independent before matching, we have 
\begin{align*}
    p_{n_1}&=\mathbb{P}(Z_{n_{1}}=z^{**}, Z_{n_{2}}=z^{*}| \mathbf{x}_{n_{1}}, \mathbf{x}_{n_{2}}, Z_{n_{1}}\wedge Z_{n_{2}}=z^{*}, Z_{n_{1}}\vee Z_{n_{2}}=z^{**})\\
    &=\frac{\mathbb{P}(Z_{n_{1}}=z^{**}, Z_{n_{2}}=z^{*}| \mathbf{x}_{n_{1}}, \mathbf{x}_{n_{2}})}{\mathbb{P}(Z_{n_{1}}=z^{**}, Z_{n_{2}}=z^{*}| \mathbf{x}_{n_{1}}, \mathbf{x}_{n_{2}})+\mathbb{P}(Z_{n_{1}}=z^{*}, Z_{n_{2}}=z^{**}| \mathbf{x}_{n_{1}}, \mathbf{x}_{n_{2}})}\\
       &=\frac{f(z^{**}| \mathbf{x}_{n_{1}}) f(z^{*}| \mathbf{x}_{n_{2}})}{f(z^{**}| \mathbf{x}_{n_{1}}) f(z^{*}| \mathbf{x}_{n_{2}})+f(z^{*}| \mathbf{x}_{n_{1}}) f(z^{**}| \mathbf{x}_{n_{2}})}, 
\end{align*}
where $f(z| \mathbf{x})$ is the generalized propensity score. We then obtain $\widehat{p}_{n_1}$ by replacing $f(z| \mathbf{x})$ with its corresponding estimate, $\widehat{f}(z| \mathbf{x})$.

\subsection{Part Two: Bias-Corrected Randomization-Based Estimation and Inference}\label{sec: bias-corrected estimation}

By incorporating a dose assignment discrepancy caliper that avoids matching individuals with highly non-random post-matching dose assignment probabilities, we can substantially reduce estimation and inference bias, as demonstrated in the simulation studies in Section~\ref{sec: simulation studies}. However, in practice, even with the caliper, non-negligible covariate imbalance may persist in some or many matched pairs after non-bipartite matching. To address this, we propose a {\it bias-corrected Neyman estimator} that leverages treatment dose and covariate information \textit{after matching} to further correct for bias due to inexact matching.

To motivate the explicit form of the proposed estimator, we first consider the following oracle bias-corrected Neyman estimator under the true generalized propensity score $f(z| \mathbf{x})$:
\begin{equation}\label{eqn: oracle bias-corrected estimator}
\widehat{\lambda}_{*}=\frac{\sum_{i=1}^{I}\left\{p_{i1}^{-1}\widetilde{Z}_{i1}(Y_{i1}-Y_{i2})+p_{i2}^{-1}\widetilde{Z}_{i2}(Y_{i2}-Y_{i1})\right\}}{\sum_{i=1}^{I}\sum_{j=1}^{2}(z^{**}_{i}-z_{i}^{*})},
\end{equation} where $p_{i1}=\mathbb{P}(Z_{i1}=z^{**}_{i}, Z_{i2}=z^{*}_{i} | \mathbf{x}_{i1}, \mathbf{x}_{i2}, Z_{i1}\wedge Z_{i2}=z^{*}_{i}, Z_{i1}\vee Z_{i2}=z^{**}_{i})=f(z^{**}_{i}| \mathbf{x}_{i1}) f(z^{*}_{i}| \mathbf{x}_{i2})$
$\{f(z^{**}_{i}| \mathbf{x}_{i1}) f(z^{*}_{i}| \mathbf{x}_{i2})+f(z^{*}_{i}| \mathbf{x}_{i1}) f(z^{**}_{i}| \mathbf{x}_{i2})\}^{-1}$ is the probability that individual $i1$ receives the higher treatment dose $z^{**}_i$, and $p_{i2}=\mathbb{P}(Z_{i1}=z^{*}_{i}, Z_{i2}=z^{**}_{i}| \mathbf{x}_{i1}, \mathbf{x}_{i2}, Z_{i1}\wedge Z_{i2}=z^{*}_{i}, Z_{i1}\vee Z_{i2}=z^{**}_{i})=1-p_{i1}$ is the probability that individual $i2$ receives the higher treatment dose $z^{**}_{i}$. Proposition~\ref{prop: unbiased} below shows that the $\widehat{\lambda}_{*}$ in \eqref{eqn: oracle bias-corrected estimator} is an unbiased estimator for the SATE $\lambda$, even under inexact matching on covariates (all the proofs in our work are in Appendix A).

\begin{Proposition}\label{prop: unbiased}
   Assume that the treatment dose of each individual is independent before matching and that there are no unobserved covariates. Then, the oracle bias-corrected Neyman estimator $\widehat{\lambda}_{*}$ in \eqref{eqn: oracle bias-corrected estimator} is unbiased for estimating the SATE, i.e., $\mathbb{E}(\widehat{\lambda}_{*}|\mathcal{F}, \mathcal{Z}) =\lambda$, even under inexact matching.
\end{Proposition}
In contrast, unless under the exact matching, the conventional Neyman estimator $\widehat{\lambda}_{N}$ that ignores the post-matching covariate imbalance should be biased for estimating the SATE $\lambda$. 
 
We are now in a position to derive a valid variance estimator for $\widehat{\lambda}_{*}$, which critically generalizes the arguments in \citet{fogarty2018mitigating} and \citet{zhang2024bridging} to the settings of inexact matching with continuous treatments. Specifically, let $\bQ$ be an arbitrary prespecified $I \times L$ matrix with $L < I$. For example, to incorporate the covariate information, we can define $\bQ = (\mathbbm{1}_{I\times 1}, \overline{\mathbf{x}}_{1}, ..., \overline{\mathbf{x}}_{K})$, where $\mathbbm{1}_{I\times 1}=(1,\dots, 1)^{T}$ and $\overline{\mathbf{x}}_{k} =((x_{11k} + x_{12k})/2,\dots, (x_{I1k} + x_{I2k})/2)^{T}$ represents the average of the $k$th covariate within each matched pair, for $k=1,\dots, K$. Let $\bH_{Q} = \bQ(\bQ^T \bQ)^{-1}\bQ^{T}$ be the projection matrix associated with $\bQ$, and let $h_{Q_{ij}}$ denote its $(i, j)$ element. Define $V_{i} = p_{i1}^{-1}\widetilde{Z}_{i1}(Y_{i1}-Y_{i2})+p_{i2}^{-1}\widetilde{Z}_{i2}(Y_{i2}-Y_{i1})$ as the contribution of the $i$th matched pair to $\widehat{\lambda}_{*}$. Now, let $y_i = V_{i}/\sqrt{1-h_{Q_{ii}}}$, define $\mathbf{y} = (y_{1},\ldots,y_{I})$, and let $\bI$ be the $I \times I$ identity matrix. We then define the variance estimator $$S_{*}^2(\bQ) = \left(2\sum_{i=1}^{I}z_{i}^{**} - z_{i}^{*}\right)^{-2}\mathbf{y}(\bI - \bH_{Q})\mathbf{y}^T,$$ the asymptotic validity of which is established in the following proposition. 

\begin{Proposition}\label{prop: conservative variance}
Assume that the treatment dose assignments are independent across matched pairs. For any $I\times L$ matrix $\bQ$ with $L<I$, we have (even under inexact matching) $$\mathbb{E}(S_{*}^2(\bQ) | \mathcal{F}, \mathcal{Z}) \geq \mathrm{Var}(\widehat{\lambda}_{*} | \mathcal{F}, \mathcal{Z}).$$
\end{Proposition}

To derive our further theoretical results, we need the following regularity conditions. These conditions can be viewed as 
generalizations of some commonly used regularity conditions in the matching literature \citep{ fogarty2018mitigating, Fogarty2023, zhao2019sensitivityvalue, zhang2022bridging, zhang2024sensitivity, zhang2024bridging} to the setting of inexact matching with continuous treatments. 

\begin{condition}[No Extreme Matched Pairs] \label{condition: asymptotic normality}
Define $\widetilde{D}_{i1} = Y_{i1}(z_{i}^{**}) - Y_{i2}(z_{i}^{*})$ and $\widetilde{D}_{i2} = Y_{i2}(z_{i}^{**}) - Y_{i1}(z_{i}^{*})$. Then, we let $V_{i}^{+} = \max\{p_{i1}^{-1}\widetilde{D}_{i1}, p_{i2}^{-1}\widetilde{D}_{i2}\}$, $V_{i}^{-}  = \min\{p_{i1}^{-1}\widetilde{D}_{i1}, p_{i2}^{-1}\widetilde{D}_{i2}\}$, and $M_i = V_{i}^{+} - V_{i}^{-}$. As $I \to \infty$, we have $$\displaystyle \underset{1 \leq i \leq I}{\text{max }} \frac{M_i^2}{\sum_{i=1}^{I}p_{i1}p_{i2}M_i^2} \to 0.$$
\end{condition} 
\begin{condition}[Bounded Fourth Moments]\label{condition: bounded fourth moments}
    Denote $q_{il}$ as the entry at the $i$th row and $l$th column of the $I\times L$ matrix $\bQ$ involved in $S^{2}_{*}(\bQ)$. There exists a constant $C < \infty$ such that for all $I$, we have $I^{-1}\sum_{i=1}^{I}M_{i}^{4} \leq C$, $I^{-1}\sum_{i=1}^{I}q_{il}^{4} \leq C$ (for all $l$), $I^{-1}\sum_{i=1}^{I}(V^{+}_{i})^{4} \leq C$, and $I^{-1}\sum_{i=1}^{I}(V^{-}_{i})^{4} \leq C$.
\end{condition}
\begin{condition}[Convergence of Finite-Population Means]\label{condition: convergence of finite-population means}
Let $\mu_{i} = \mathbb{E}(V_i |\mathcal{F}, \mathcal{Z}) = \widetilde{D}_{i1} + \widetilde{D}_{i2}$ and $\nu_i^2 = \mathrm{Var}(V_i |\mathcal{F}, \mathcal{Z}) = p_{i1}p_{i2}(p_{i1}^{-1}\widetilde{D}_{i1} - p_{i2}^{-1}\widetilde{D}_{i2})^2$. As $I \to \infty$, we have 1) $I^{-1}\sum_{i=1}^{I}\mu_{i}^2$ converges to a finite positive value, 2) $I^{-1}\sum_{i=1}^{I}\nu_{i}^2$ converge to a finite positive value, 3) for all $l = 1,...,L$, $I^{-1}\sum_{i=1}^{I} \mu_i q_{il}$ converge to a finite value, and 4) $I^{-1}\bQ^{T}\bQ$ converges to a finite, invertible $L \times L$ matrix $\widetilde{\bQ}$.
\end{condition}

Condition~\ref{condition: asymptotic normality} ensures that no single matched pair contributes disproportionately to the average treatment effect estimation, so the total contribution from the entire dataset dominates that of any individual pair. Such a regularity condition has been widely used in the matching literature  \citep{zhao2019sensitivityvalue, zhang2022bridging, Fogarty2023, zhang2024sensitivity}. A set of sufficient conditions to satisfy Condition~\ref{condition: asymptotic normality} includes: 1) $M_i^2$ is bounded for each $i$, and 2) there exists an $\epsilon > 0$ such that $\epsilon \leq p_{i1} \leq 1-\epsilon$ for each matched pair $i$. Additionally, Condition~\ref{condition: bounded fourth moments}, which imposes bounded fourth moments, and Condition~\ref{condition: convergence of finite-population means}, which assumes the convergence of finite-population means, are analogous to regularity conditions commonly adopted in matching or stratification with binary treatments \citep{fogarty2018mitigating, Fogarty2023}, while being tailored to the continuous treatment setting.

Then, as shown in Theorem~\ref{thm: valid confidence interval} below, under the true generalized propensity scores, an asymptotically valid confidence interval for the SATE $\lambda$ can be constructed based on $\widehat{\lambda}_{*}$ and the corresponding variance estimator $S_{*}^{2}(\bQ)$. 

\begin{theorem}\label{thm: valid confidence interval}
Assuming independence of dose assignments across matched pairs and no unmeasured confounding, as well as Conditions 1--3 stated above, the coverage rate of confidence interval $(\widehat{\lambda}_{*}-\Phi^{-1}(1-\alpha/2) S_{*}(\bQ), \widehat{\lambda}_{*}+\Phi^{-1}(1-\alpha/2) S_{*}(\bQ))$ for the SATE $\lambda$ is asymptotically no less than $100(1-\alpha)\%$, where $\Phi$ is the cumulative distribution function of the standard normal distribution and $\alpha\in (0,0.5)$ is some prespecified level $\alpha$. 
\end{theorem}
To our knowledge, Theorem~\ref{thm: valid confidence interval} gives the first asymptotically valid confidence interval for the SATE under inexact matching beyond the binary treatment case. The core idea of the proof is to extend the arguments in \cite{fogarty2018mitigating, Fogarty2023} from the exact matching and binary treatment case to the inexact matching and continuous treatment case.

In practice, the true generalized propensity score $f(z| \mathbf{x})$ in $\widehat{\lambda}_{*}$ and $S_{*}^{2}(\bQ)$ is unknown. A natural and commonly adopted strategy is to plug an estimated generalized propensity score $\widehat{f}(z| \mathbf{x})$ into the $\widehat{\lambda}_{*}$, leading to the following bias-corrected Neyman estimator $\widehat{\lambda}$:
\begin{equation}
\widehat{\lambda}=\frac{\sum_{i=1}^{I}\left\{\widehat{p}_{i1}^{-1}\widetilde{Z}_{i1} (Y_{i1}-Y_{i2})+\widehat{p}_{i2}^{-1}\widetilde{Z}_{i2} (Y_{i2}-Y_{i1})\right\}}{\sum_{i=1}^{I}\sum_{j=1}^{2}(z^{**}_{i}-z_{i}^{*})},
\label{eqn: vanilla bias-corrected estimator}
\end{equation}
where $\widehat{p}_{i1}=\widehat{f}(z^{**}_{i}|\mathbf{x}_{i1}) \widehat{f}(z^{*}_{i}|\mathbf{x}_{i2})\{\widehat{f}(z^{**}_{i}| \mathbf{x}_{i1}) \widehat{f}(z^{*}_{i}| \mathbf{x}_{i2})+\widehat{f}(z^{*}_{i}| \mathbf{x}_{i1}) \widehat{f}(z^{**}_{i}| \mathbf{x}_{i2})\}^{-1}$ and $\widehat{p}_{i2}=1-\widehat{p}_{i1}$. Similarly, we can use this plug-in strategy to derive the corresponding bias-corrected variance estimator $S^{2}(\bQ)$, which is obtained by replacing each $p_{ij}$ in $S_{*}^{2}(\bQ)$ with estimate $\widehat{p}_{ij}$. 

\begin{remark}\label{rem: regularization}
For $\widehat{p}_{ij}$ close to $1$ or $0$, either $\widehat{p}_{i1}^{-1}$ or $\widehat{p}_{i2}^{-1}$ is large, which may introduce high variance and large finite-sample bias in the bias-corrected Neyman estimator $\widehat{\lambda}$. Incorporating the dose assignment discrepancy caliper into the non-bipartite matching procedure (as described in Section~\ref{sec: calipered matching}) can effectively avoid this issue. If there still remain some matched pairs $i$ with $\widehat{p}_{ij}$ close to $1$ or $0$, we can replace $\widehat{p}_{ij}$ by a regularized version $\widehat{p}_{ij} = \mathbbm{1}\{\delta \leq \widehat{p}_{ij} \leq 1-\delta\} + \delta  \mathbbm{1}\{\widehat{p}_{ij} < \delta\}+ (1-\delta) \mathbbm{1}\{\widehat{p}_{ij} > 1-\delta\}$, 
with prespecified $\delta \in (0, 0.5)$ denoting the regularization threshold. This regularization strategy follows a similar spirit to the trimming or truncation strategy commonly used in the weighting literature \citep{crump2009dealing, ma2020robust}. Throughout the simulation studies and data application, we use the regularized bias-corrected Neyman estimator, setting the regularization threshold $\delta$ as the commonly used value $0.1$ \citep{crump2009dealing, sturmer2021propensity}.
\end{remark}

\section{Simulation Studies}\label{sec: simulation studies}

\subsection{Simulation Settings}

In this section, we assess the performances of the conventional and bias-corrected Neyman estimators by comparing their empirical bias. For inference, we compare the conventional Neyman-type confidence interval and the bias-corrected confidence interval in terms of the mean confidence interval length (MCIL) and the coverage rate (CR). Two matching schemes are considered: 
\begin{itemize}
    \item {Non-bipartite matching without caliper:} This strategy forms matched pairs by minimizing the total Mahalanobis distance of observed covariates across all pairs \citep{lu2001matching, lu2011optimal, baiocchi2010building, rosenbaum2020modern}. It is implemented via the widely used \texttt{nbpMatching} package \citep{lu2011optimal} and is considered the standard practice. 
    \item {Non-bipartite matching with a dose assignment discrepancy caliper:} We implement our dose assignment discrepancy caliper together with the non-bipartite matching procedure by using the proposed distance metric in \eqref{eqn: distance metric with penalty}. To estimate the generalized propensity score $f(z| \mathbf{x})$ for computing $\widehat{p}_{n}$ in \eqref{eqn: distance metric with penalty}, we consider three methods: 
    \begin{enumerate}
        \item[(i)] Conditional density estimation via Lindsey's method and boosting, as described in \citet{gao2022lincde}, implemented in \texttt{R} package \texttt{linCDE}.
        \item[(ii)] A random forest-based conditional density estimator from \citet{pospisil2018rfcde}, available in the \texttt{RFCDE} package.
        \item[(iii)] A ``model-based" method assumes that the conditional density of treatment $Z$ given covariates $\mathbf{X}$ follows $Z|\mathbf{X} \sim \mathbb{E}(Z| \mathbf{X}) + \epsilon$, where $\epsilon \sim N(0, \sigma^2)$. That is, $f(Z| \mathbf{X}) = \sigma^{-1}\psi((Z - \mathbb{E}(Z| \mathbf{X}))/\sigma)$, where $\psi(\cdot)$ is the probability density function of the standard normal distribution. In our simulations, we estimate $\mathbb{E}(Z|\mathbf{X})$ using multivariate adaptive regression splines \citep{friedman1991multivariate}. 
    \end{enumerate}  We set $\xi=0.1$ and $M=\infty$ or a sufficiently large number. Although we primarily focus on methods (i)-(iii) to illustrate our framework, it is flexible enough to accommodate any generalized propensity score estimators, i.e. any conditional density estimation methods.
\end{itemize}

In addition to the above two matching schemes, we also consider the following two estimation and inference methods: 
\begin{itemize}
    \item Estimation and inference based on the conventional Neyman estimator $\widehat{\lambda}_{N}$, considered as the common practice in matched observational studies with continuous treatments \citep{baiocchi2010building, rosenbaum2020design, fogarty2021biased, heng2023instrumental}.

    \item Bias-corrected estimation and inference based on the bias-corrected Neyman estimator $\widehat{\lambda}$. For implementation, we use the three methods mentioned above to estimate the generalized propensity score $f(z| \mathbf{x})$: \texttt{linCDE},  \texttt{RFCDE}, and the ``model-based" method.
\end{itemize}

Finally, we consider two data generation settings with the sample size $N=2I=1000$: 
\begin{itemize}
    \item Setting 1:  For each $n$ (before matching), we generate covariates $X_{n1},...,X_{n5}\overset{\mathrm{i.i.d.}}{\sim} N(0, 1)$. Treatment doses $Z_{n}$ are generated as $Z_{n} = X_{n1} + X_{n2}^2 + |X_{n3}X_{n4}| + \mathbbm{1}\{X_{n4} > 0\} + \log(1 + |X_{n5}|) + \epsilon^{Z}_{n}$, where $\epsilon^{Z}_{n} \overset{\mathrm{i.i.d.}}{\sim} N(0, 1)$. Potential outcomes $Y_{n}(z)$ are generated as $Y_{n}(z) = z + 0.3  X_{n1}z + 0.2X^{3}_{n3}z + \exp\{|X_{n2} - X_{n4}|\} - \sin(X_{n5}) + \epsilon^{Y}_{n}$, where $\epsilon^{Y}_{n}\overset{\mathrm{i.i.d.}}{\sim} N(0, 3)$.
    \item Setting 2: We generate independent covariates with the following distributions: $X_{n1} \overset{\mathrm{i.i.d.}}{\sim} \text{Ber}(0.5), X_{n2} \overset{\mathrm{i.i.d.}}{\sim} \text{Bin}(10, 0.75), X_{n3} \overset{\mathrm{i.i.d.}}{\sim} \text{Poi}(1.5), X_{n4} \overset{\mathrm{i.i.d.}}{\sim} \text{Unif}(0, 3), X_{n5} \overset{\mathrm{i.i.d.}}{\sim} \text{Unif}(-1, 1)$, and $X_{n6} \overset{\mathrm{i.i.d.}}{\sim} \text{Unif}(-5, 5)$. Treatment doses $Z_{n}$ are generated as $Z_{n} = 5 \mathbbm{1}\{X_{n1} = 1, X_{n2} > 5\} - X_{n1} + X_{n3} + \sin^{2}(X_{n5}) + 2 \log(1 + X_{n4}) + 2\exp(-|X_{n6}|) + \epsilon^{Z}_{n}$, where $\epsilon^{Z}_{n} \overset{\mathrm{i.i.d.}}{\sim} N(0, 1)$. Potential outcomes $Y_{n}(z) = z + (0.7  X_{n6} + 2X^{3}_{n5})z + X_{n3}\{(2X_{n4} + 1)\mathbbm{1}\{X_{n1}=1\} + \mathbbm{1}\{X_{n1}=0\}\} +0.5X_{n4}|X_{n5}| +X_{n2} + \epsilon^{Y}_{n}$, where $\epsilon^{Y}_{n}\overset{\mathrm{i.i.d.}}{\sim} N(0, 3)$.
\end{itemize}
In both settings, when generating the matched datasets, we only keep the matched datasets that meet the commonly used covariate balance criteria -- absolute standardized mean differences of all covariates between paired individuals are below $0.2$ \citep{rosenbaum2020design, zhang2022bridging}. As shown in Table S1 in Appendix B, the average absolute standardized mean differences for all covariates across the $1,000$ simulated matched datasets are consistently below, and typically much lower than, $0.08$. That is, based on the standard practice, these matched datasets would have been viewed as sufficiently balanced in covariates, and researchers would adopt the classical randomization-based estimation and inference without bias correction for inexact matching \citep{rosenbaum2020design, zhang2022bridging}.

\subsection{Simulation Results}
As shown in Table~\ref{tab: sim1est}, incorporating the proposed dose assignment discrepancy caliper during matching usually reduces the estimation bias of the conventional Neyman estimator and that of the bias-corrected Neyman estimator. In addition, the proposed bias-corrected Neyman estimator generally achieves a substantial reduction in estimation bias compared to the conventional Neyman estimator, regardless of whether the proposed caliper is applied. Finally, the choice of estimation of the generalized propensity score may influence the performance of both the dose-assignment discrepancy caliper and the bias-corrected Neyman estimator, with the optimal choice depending on the underlying data-generating process.

\begin{table}[H]
    \centering
    \caption{The average estimation bias for the combinations of different matching strategies (optimal non-bipartite matching without or with the dose assignment discrepancy caliper), estimation methods (conventional or bias-corrected Neyman estimator), and two data generation settings. When applying the proposed caliper and bias-corrected estimator, three methods for estimating generalized propensity scores are ``\texttt{linCDE}," ``\texttt{RFCDE}," and the model-based method.}
    \setlength\extrarowheight{-6pt}
    \addtolength{\tabcolsep}{-0.1em}
        \begin{tabular}{ccccc}\\
         \toprule
         & & \multicolumn{3}{c}{Bias-Corrected} \\
         \cmidrule(rl){3-5}
         \multirow{-2}{*}{Setting 1} & \multirow{-2}{*}{Conventional} & \texttt{linCDE} & \texttt{RFCDE} & Model-Based \\
          \hline
         No Caliper  & 0.904 & 0.692 & 0.143 & 0.300 \\
      Caliper (\texttt{linCDE})  & 0.875 & 0.718 & 0.132 & 0.288 \\
       Caliper (\texttt{RFCDE})   & 0.801 & 0.759 & 0.132 & 0.276 \\
        Caliper (Model-Based)  & 0.328 & 0.272 & 0.240 & 0.018 \\
          \bottomrule
    \end{tabular}
    \label{tab: sim1est}
        \vspace{8mm}
    \begin{tabular}{ccccc}
         & & \multicolumn{3}{c}{Bias-Corrected} \\
                  \cmidrule(rl){3-5}
         \multirow{-2}{*}{Setting 2} & \multirow{-2}{*}{Conventional} & \texttt{linCDE} & \texttt{RFCDE} & Model-Based \\
          \hline
         No Caliper  & 0.268 & 0.001 & 0.189 & 0.031 \\
         Caliper (\texttt{linCDE})  & 0.221 & 0.002 & 0.207 & 0.036 \\
        Caliper (\texttt{RFCDE}) & 0.187 & 0.023 & 0.218 & 0.063 \\
        Caliper (Model-Based)  & 0.179 & 0.011 & 0.225 & 0.027 \\
          \bottomrule
    \end{tabular}
    \label{tab: sim2est}
\end{table}

The inference results in Table~\ref{tab: sim1inference} indicate that the proposed bias-corrected confidence interval, which is derived from the bias-corrected Neyman estimator and its corresponding bias-corrected variance estimator, usually achieves a higher coverage rate compared to the SATE confidence interval based on the conventional Neyman estimator. Additionally, applying the proposed dose assignment discrepancy caliper generally increases variance estimates across all estimators but also improves overall coverage rates. Also, the length and coverage of the bias-corrected confidence interval depend on the underlying data-generating process and the generalized propensity score estimators used.
 
\begin{table}[H]
    \centering
    \caption{The average $95\%$ confidence interval length (MCIL) and coverage rate (CR) for the combinations of different matching strategies (optimal non-bipartite matching without or with the dose assignment discrepancy caliper), estimation methods (conventional or bias-corrected Neyman estimator), and two data generation settings. When applying the proposed caliper and bias-corrected estimator, three methods for estimating generalized propensity scores are ``\texttt{linCDE}," ``\texttt{RFCDE}," and the model-based method.}
    \setlength\extrarowheight{-6pt}
    \begin{tabular}{cc|cccc}\\
    \toprule
           \multirow{2}{*}{Setting 1} & & \multirow{2}{*}{Conventional} & \multicolumn{3}{c}{Bias-Corrected}  \\
                    \cmidrule(rl){4-6}

           &&& \texttt{linCDE} &  \texttt{RFCDE} & Model-Based\\
          \cmidrule{1-6}
          & MCIL &  1.224 & 1.129 & 0.754 &1.026 \\
          \multirow{-2}{*}{No Caliper} & CR & 0.146 & 0.370 & 0.841 & 0.740 \\
          \cmidrule{1-6}
      & MCIL  & 1.262 & 1.287 & 0.782 & 1.084 \\
         \multirow{-2}{*}{Caliper (\texttt{linCDE})} &CR & 0.191 & 0.380 & 0.842 & 0.766 \\
          \cmidrule{1-6}
              & MCIL  & 1.793 & 1.918 & 1.185 & 1.688 \\
       \multirow{-2}{*}{Caliper (\texttt{RFCDE})} & CR & 0.561 & 0.648 & 0.862 & 0.850 \\
          \cmidrule{1-6}
          & MCIL  & 1.568 & 1.711 & 1.035 & 1.645 \\
         \multirow{-2}{*}{Caliper (Model-Based)}  & CR & 0.771 & 0.816 & 0.698 & 0.851 \\
    \end{tabular}
    \label{tab: sim1inference}
    \vspace{8mm}
        \begin{tabular}{cc|cccc}
         \toprule
          \multirow{2}{*}{Setting 2} & &  \multirow{2}{*}{Conventional} & \multicolumn{3}{c}{Bias-Corrected}  \\
                              \cmidrule(rl){4-6}

          &&& \texttt{linCDE}  & \texttt{RFCDE} & Model-Based \\
          \cmidrule{1-6}
          & MCIL  & 1.064 & 1.052 & 0.733 & 1.161 \\
          \multirow{-2}{*}{No Caliper} & CR & 0.867 & 0.969 & 0.846 & 0.965 \\
          \cmidrule{1-6}
         & MCIL  & 1.141 & 1.158 & 0.795 & 1.295 \\
        \multirow{-2}{*}{Caliper (\texttt{linCDE})}  & CR & 0.896 & 0.974 & 0.845 & 0.976 \\
          \cmidrule{1-6}
          & MCIL &  1.165 & 1.212 & 0.823 & 1.346 \\
          \multirow{-2}{*}{Caliper (\texttt{RFCDE})} & CR & 0.921 & 0.966 & 0.834 & 0.961 \\
          \cmidrule{1-6}
          & MCIL  & 1.252 & 1.312 & 0.881 & 1.404 \\
        \multirow{-2}{*}{Caliper (Model-Based)} & CR & 0.934 & 0.966 & 0.840 & 0.965 \\
          \bottomrule
    \end{tabular}
    \label{tab: sim2inference}
\end{table}

Tables~\ref{tab: sim1est} and~\ref{tab: sim1inference} convey a unified message: the common practice of constructing non-bipartite matching without a carefully designed caliper, ignoring post-matching covariate imbalance, and relying solely on the conventional Neyman estimator can introduce severe bias in SATE estimation and inference. This holds even when post-matching covariate balance meets commonly used empirical criteria. The severity of bias depends on the outcome- and treatment dose-generating processes. The proposed bias mitigation framework has strong potential to substantially reduce estimation bias and improve confidence interval coverage. 

\section{Data Application: Re-Analyzing the Effects of Social Distancing Practice On COVID-19 Case Counts}\label{sec: real data application}

As discussed in Section~\ref{subsec: motivating example}, using the conventional Neyman estimator to analyze the matched dataset on social distancing and COVID-19 case counts suggests that social distancing may substantially \textit{increase} COVID-19 cases among U.S. counties \citep{zhang2023social}. We re-examine this conclusion using our proposed bias mitigation framework. Specifically, we apply the proposed bias-corrected Neyman estimator (setting $\delta=0.1$) to construct a point estimate and $95\%$ confidence interval for the SATE, first on the original $1,211$ matched pairs of U.S. counties from \citet{zhang2023social} and then on a newly formed $1,211$ matched pairs of U.S. counties using optimal non-bipartite matching with the proposed dose assignment discrepancy caliper (setting $\xi = 0.1$).

\begin{figure}[ht]
\caption{Left: Distribution of estimated treatment assignment probabilities for all potential matched pairs. Right: Distribution of estimated treatment assignment probabilities for matched pairs from \cite{zhang2023social}.}
\vspace{0.2cm}
\centering
\begin{subfigure}
  \centering
  \includegraphics[width=.45\linewidth]{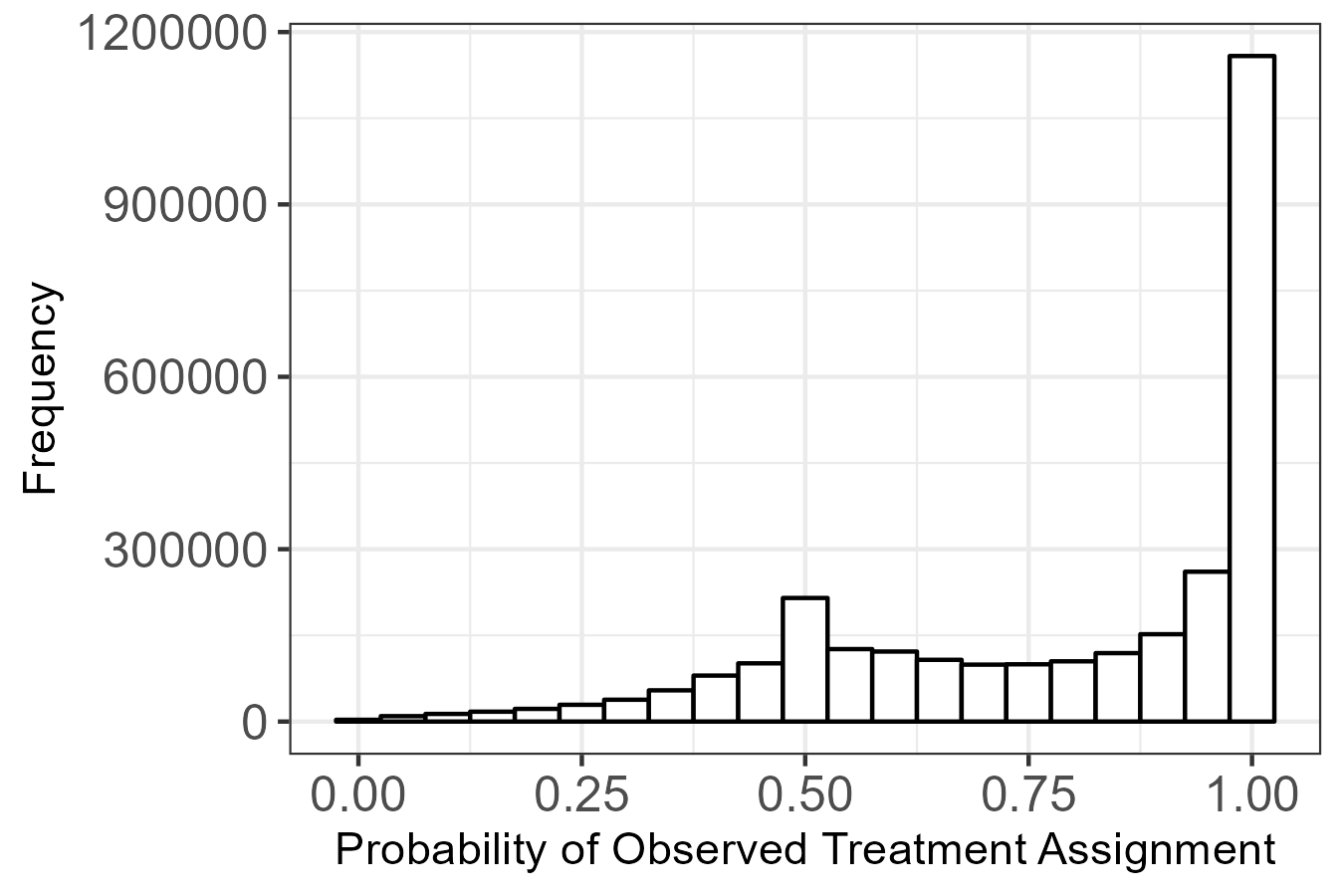}
  \label{fig:sub1}
\end{subfigure}
\begin{subfigure}
  \centering
  \includegraphics[width=.45\linewidth]{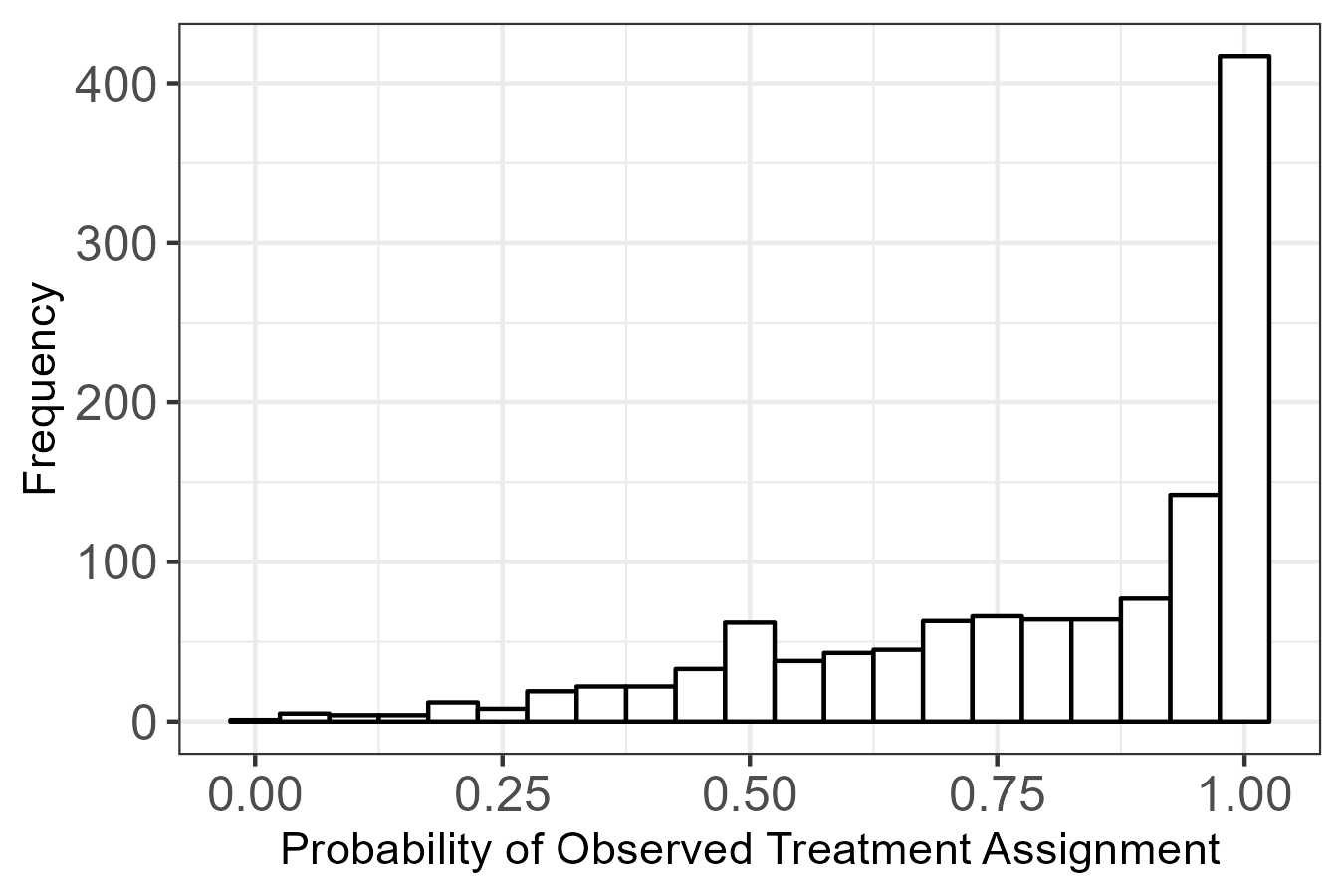}
  \label{fig:sub2}
\end{subfigure}
\label{fig:p_dist}
\end{figure}

We use the $\texttt{R}$ package \texttt{linCDE} to stimate the generalized propensity scores involved in our methods. As shown in Figure~\ref{fig:p_dist}, the estimated post-matching treatment dose assignment probabilities for the original matched dataset from \cite{zhang2023social} heavily deviate from $0.5$. Therefore, we expect our method’s results to differ substantially from those obtained using the conventional Neyman estimator on the original matched dataset.

Table~\ref{tab: data analysis} shows SATE estimates and confidence intervals for the effect of social distancing, measured by county-level percentages of reduction in total distance traveled, on cumulative COVID-19 cases during the follow-up period. These results are based on four combinations: the conventional Neyman estimator or the bias-corrected Neyman estimator applied to either the original matched pairs \citep{zhang2023social} or the newly formed matched pairs using the proposed dose assignment discrepancy caliper. Applying the conventional Neyman estimator to the original matched pairs yields a SATE estimate of $-1041.478$ with $95\%$ confidence interval $(-1399.777, -683.180)$, as reported in Section~\ref{subsec: motivating example}. This implies that,  averaging over the matched counties, reducing total distance traveled by $50\%$ during the re-opening phase (April 27-June 28, 2020), in comparison with the pre-pandemic baseline, is estimated to \textit{increase} cumulative COVID-19 cases by $(-0.5)(-1041.478) = 520.739$ per $100,000$ people during the five-week follow-up (June 19-August 2, 2020). This seemingly counterintuitive result, suggesting that more stringent social distancing increases COVID-19 cases, contradicts established public health understanding.

In contrast, applying the bias-corrected Neyman estimator to the newly formed matched pairs yields a SATE estimate of $361.640$, overturning the earlier conclusion and instead suggesting that stricter social distancing helps prevent COVID-19 transmission. While the $95\%$ confidence interval does not rule out the null effect, likely due to the larger variance associated with the caliper (as suggested by Table~\ref{tab: sim1inference}), further modification could improve the efficiency of the bias-corrected Neyman estimator, for example, through an enhanced covariate adjustment in variance estimation $S^{2}(\bQ)$. Moreover, Table~\ref{tab: data analysis} shows that the dose assignment discrepancy caliper and the bias-corrected Neyman estimator, whether implemented individually or together, can substantially shift the SATE estimate and confidence interval compared with these reported by the conventional Neyman estimator applied to the original matched dataset, offering different and more interpretable scientific insights.

\begin{table}[h]
\caption{Point estimates and $95\%$ confidence intervals for the effect of social distancing on cumulative COVID-19 case counts over a five-week follow-up period. Results are shown for four combinations: the conventional Neyman estimator or the bias-corrected Neyman estimator applied to either the original matched pairs from \citet{zhang2023social} or the newly formed $1,211$ matched pairs using the proposed dose assignment discrepancy caliper.}
\vspace{0.2cm}
\centering
\begin{tabular}[ht]{ccrr}
\toprule
Caliper & Estimator & Estimate & 95\% Confidence Interval \\
\midrule
& Conventional & $-1041.478$ & $(-1399.777, -683.180)$ \\
\multirow{-2}{*}{No} & Bias-Corrected & $-474.556$ & $(-767.866, -181.247)$ \\
\midrule
& Conventional & $-136.968$  & $(-673.094, 399.158)$ \\
\multirow{-2}{*}{Yes} & Bias-Corrected &  $361.640$  & $(-98.059, 821.340)$ \\
\hline
\end{tabular}
\label{tab: data analysis}
\end{table}

\section{Conclusion and Discussion}\label{sec: discussion}

In this work, we introduce a bias mitigation framework for inexactly matched observational studies with continuous treatments, proposing adjustments in matching, estimation, and inference stages. Compared to common practice, our framework better leverages post-matching covariate imbalance information to reduce estimation and inference bias effectively. It naturally extends beyond the continuous treatment case and can be applied to ordinal treatments as well.

The numerical studies and re-analysis of social distancing effects on COVID-19 case counts convey two key messages. First, even when a matched dataset meets commonly used covariate balance criteria, remaining covariate imbalance and discrepancies in post-matching treatment dose assignment probabilities can introduce substantial bias in downstream treatment effect estimation and inference. This underscores the need to improve non-bipartite matching by incorporating a caliper tailored to downstream analysis, as well as adjusting for inexact matching bias at the estimation and inference stages. Second, as shown in the numerical studies, the extent of improvement using our framework depends on the choice of generalized propensity score estimation. It is crucial to study the theoretical properties of different conditional density estimation methods in our context and determine which is best suited for integration into our framework across various data-generating processes. We leave this for future work.

\section*{Acknowledgements}

The authors thank Rebecca Betensky, Colin Fogarty, Debashis Ghosh, Hyunseung Kang, Bo Zhang, Jeffrey Zhang, and Jianan Zhu for the helpful comments and discussions. The work of Siyu Heng was supported in part by the National Institute on Drug Abuse of the National Institutes of Health (NIH) under Award Number R21DA060433, as well as an NYU Research Catalyst Prize. Wen Zhou was supported partially by NIH R01GM144961, NIH R01GM157600, and NSF-IOS 1922701. The content is solely the responsibility of the authors and does not necessarily represent the official views of the NIH or NSF.


\section*{Supplementary Materials}
Web Appendices A--C are available on the Biometrics website at Oxford Academic. Reproduction {\tt R} code for simulations and data examples is available at \url{https://github.com/AnthonyFrazierCSU/Mitigating-Bias-Matched-Observational-Studies}. An open-source \texttt{R} package \texttt{nbpInference} implementing our framework is available at \url{https://github.com/AnthonyFrazierCSU/nbpInference}.

\bibliographystyle{apalike}

\end{document}